\documentclass[USenglish,runningheads]{llncs}[2018/03/10]

\usepackage[USenglish]{babel}

\usepackage[utf8]{inputenc} 

\usepackage{graphicx}
\usepackage{subfig}

\usepackage{upquote}

\usepackage{booktabs}

\usepackage{paralist}

\usepackage{csquotes}

\usepackage{url}

\usepackage{listings}
\renewcommand{\lstlistingname}{List.}
\let\counterwithout\relax

\usepackage{chngcntr}
\AtBeginDocument{\counterwithout{lstlisting}{section}}

\usepackage[%
  square,   comma,
   numbers,
   sort&compress,
]{natbib}

\usepackage{hyperref}
\hypersetup{hidelinks, colorlinks=true, allcolors=black, pdfstartview=,
pdftitle={JArena: Partitioned Shared Memory for NUMA-awareness of
Multithreaded Scientific Appplications},breaklinks=true}
\usepackage[all]{hypcap}

\usepackage[capitalise,nameinlink]{cleveref}
\crefname{section}{Sect.}{Sect.}
\Crefname{section}{Section}{Sections}
\crefname{listing}{\lstlistingname}{\lstlistingname}
\Crefname{listing}{Listing}{Listings}

\begin{document}

\title{JArena: Partitioned Shared Memory for NUMA-awareness in Multi-threaded
  Scientific Applications}
\titlerunning{JArena: Partitioned shared memory for NUMA-awareness}


\author{%
  Zhang Yang\inst{1,2,3} \and Aiqing Zhang \inst{1,2,3} \and
  Zeyao Mo \inst{1,2}
}
\authorrunning{Zhang Yang et al.}

\institute{%
  Institute of Applied Physics and Computational Mathematics \\
  \email{yang\_zhang,zhang\_aiqing,zeyao\_mo@iapcm.ac.cn} \and CAEP Software Center for High Performance Numerical
  Simulation \\ \and Key Laboratory of Computational Physics
}

\maketitle

\begin{abstract}
  The distributed shared memory (DSM) architecture is widely used in today's
  computer design to mitigate the ever-widening processing-memory gap, and
  inevitably exhibits non-uniform memory access (NUMA) to shared-memory parallel
  applications. Failure to achieve full NUMA-awareness can significantly
  downgrade application performance, especially on today's manycore platforms
  with tens to hundreds of cores. Yet traditional approaches such as first-touch
  and memory policy fail short in either false page-sharing, fragmentation, or
  ease-of-use.  In this paper, we propose a partitioned shared memory approach
  which allows multi-threaded applications to achieve full NUMA-awareness with
  only minor code changes and develop a companying NUMA-aware heap manager which
  eliminates false page-sharing and minimizes fragmentation. Experiments on a
  256-core cc-NUMA computing node show that the proposed approach achieves true
  NUMA-awareness and improves the performance of typical multi-threaded scientific
  applications up to 4.3 folds with the increased use of cores.
\end{abstract}


%

\section{Introduction}%
\label{sec:intro}

The distributed shared-memory (DSM) architecture (a.k.a. NUMA) is becoming
increasingly popular in computer design, especially in the manycore era. The
Kinghts Landing generation of Intel Xeon Phi processors introduces sub-NUMA
cluster modes for best possible performance, and the EPYC family of AMD
processors uses NUMA as the major approach to maintain high memory bandwidth.
ARM-based manycore processors such as Phytium 2000+ and Cavium ThunderX2 also
incorporate NUMA either intra or inter sockets. Furthermore, most supercomputers
configure their computing nodes as multiple NUMA domains.

Failure to achieve full NUMA-awareness can significantly downgrade the
performance of multi-threaded parallel applications, which can go as large as
200\% for memory-bounded applications on multi-socket cc-NUMA systems
(\cite{Majo2013M}). Thus it is crucial for applications to achieve full
NUMA-awareness to be performant. However, although special care has been taken
in areas such as the MPI runtime (\cite{Li2013N}) and graph analytics
(\cite{Zhang2015N}), the area of scientific applications still largely rely
almost solely on OS-based approaches such as first-touch and memory policy. We
argue in this paper that these approaches are far from optimal, in the sense of
both performance and ease-of-use. To be short, these approaches suffer from both
false page-sharing and fragmentation, due to the page-based memory management
mechanism. Working around these shortcomings would require an
application-specific redesign of both data structures and memory management
schemes (\cite{Majo2013M}), which would soon become a burden for application
developers.

To address the above problems, we propose a partitioned shared memory approach
in this paper. Our approach consists of an application-level conceptual
abstraction of partitioned shared memory and a companying NUMA-aware
multi-threaded heap manager named JArena. The abstraction provides applications
with a thread-partitioned view of shared memory and allows them to take full
control of on which thread one memory block resides, thus enable them to achieve
full NUMA-awareness with minor code changes. JArena in turn manages these memory
blocks to make sure no page-sharing happens across NUMA nodes and reduces
fragmentation with advanced segregated storage schemes, thus ensures
NUMA-awareness. Compared to first-touch, our approach can improve the
performance of real-world scientific applications up to 4.3 folds on a 256-core
cc-NUMA system, with only minor code changes. In summary, this paper makes the
following contributions:
\begin{enumerate}
  \item{\emph{Identify pitfalls of OS-provided NUMA-awareness approaches}}: We
        use both theoretical analysis and experiments to show that first-touch
        and memory policy would suffer from false page-sharing and
        fragmentation.
  \item{\emph{Propose a partitioned shared-memory NUMA-awareness approach}}:
        The proposed approach is conceptually simple yet adequate for full
        NUMA-awareness of multi-threaded scientific applications.
  \item{\emph{Design a NUMA-aware multi-threaded heap manager}}: We
        design and implement a general-purpose NUMA-aware multi-threaded heap manager
        which eliminates false page-sharing and reduces fragmentation.
\end{enumerate}

The rest of this paper is structured as follows: We analyze the pitfalls of
OS-based NUMA-awareness approaches in \cref{sec:background} and present the
proposed partitioned shared memory approach in \cref{sec:psm} and
JArena heap manager in \cref{sec:jarena-algorithms}. Experimental results are shown
in \cref{sec:exp} and related works are presented in
\cref{sec:related-works}. Finally comes the conclusions and the outlooks in
\cref{sec:conclusion}.

%

\section{Pitfalls of OS-based NUMA-awareness Approaches}%
\label{sec:background}

\begin{figure}[htbp]
  \vspace{-0.5cm}
  \centering
  \includegraphics[width=0.8\textwidth]{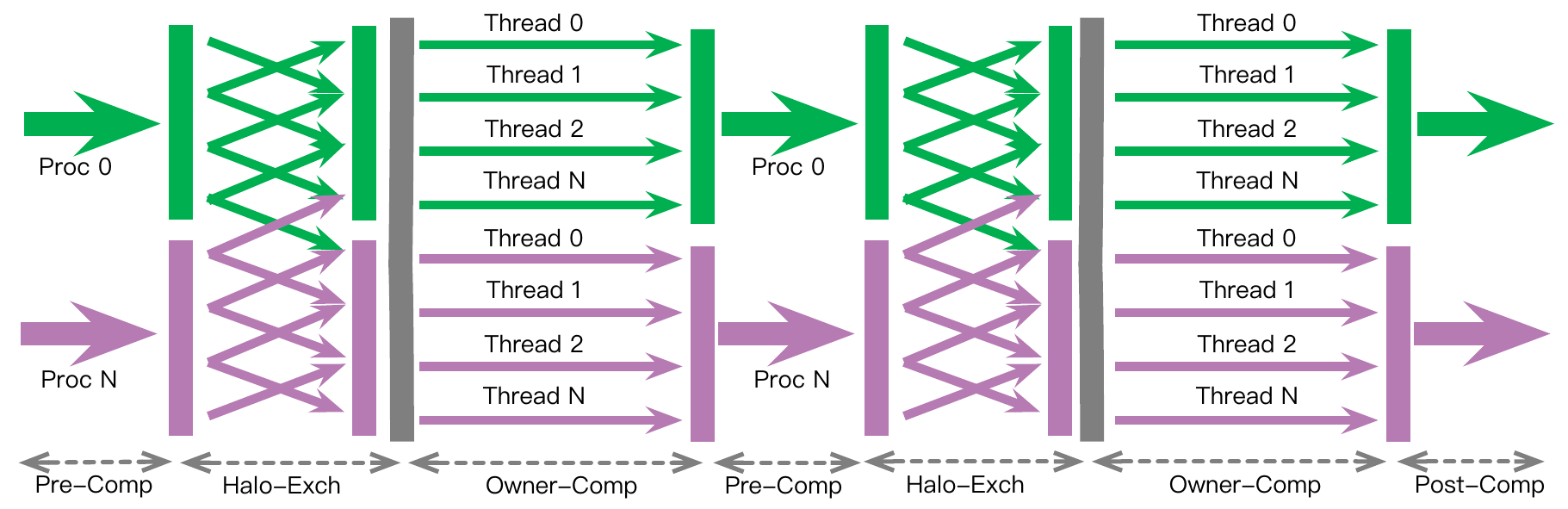}
  \caption{Illustration of BSP execution of MPI+Thread parallel
    applications.}%
  \label{fig:hybrid_bsp_execution}
  \vspace{-0.5cm}
\end{figure}

In this section, we analyze the pitfalls of OS-based NUMA-awareness approaches.
Parallel scientific applications are often built upon the MPI+Thread model and
exhibit execution patterns similar to \cref{fig:hybrid_bsp_execution}. The
application executes in locksteps and each lockstep often consists of a
inter-thread exchange phase and an owner-computation phase. The combination of
these two phases and multiple locksteps can make the memory access pattern
highly irregular as those in multi-block applications\cite{Guo2012A} and
sweepings\cite{Mo2014}.

The recommended approach for scientific applications to achieve NUMA-awareness
is the ``first-touch'' approach, where the application arranges the owner of an
data block to write to the pages first\cite{Hager2010}. When first-touch is
inadequate, explicit memory policy mechanisms such as
``numactl''\cite{Kleen2005} are used to instruct the OS where a range of pages
shall be placed. Behind the scene, the OS may introduce statistics-based
automatic page migration to further ``optimize'' locality
(\cite{Goglin2009E,Gorman2012A}). These OS-based NUMA-awareness approaches
suffer from both performance and ease-of-use. 

Firstly, since OS-based approaches can only work with fix-sized pages and
applications usually work with variable-sized blocks, fragmentation can become
significant. For typical single-thread patch sizes in structured-grid scientific
applications such as Uintah\cite{Berzins2016E} and JASMIN\cite{Mo2010}, one can
computed the fragmentation rate on different page sizes as in
\cref{tbl:memory-fragmentation}. Even with 4K page, one can waste 21.9\% the
memory in extreme cases. And for huge pages such as the 2MB page on x86\_64, the
fragmentation is unacceptably high.
\begin{table}[htpb]
  \vspace{-0.5cm}
  \centering
  \caption{Calculated fragmentation of different patch sizes w.r.t.\ different
  page sizes.}%
  \label{tbl:memory-fragmentation}
  \begin{tabular*}{\textwidth}{@{\extracolsep{\fill}}rrrrr}
    \toprule
    Patch Size & 20x20 & 50x50 & 10x10x10 & 30x30x30 \\
    \midrule
    Data size &3200B & 4000B &  8000B & 216000B \\
    \midrule
    4K Page  & 21.9\% & 2.3\%  & 2.3\%  & 0.5\%  \\
    64K Page & 95.1\% & 69.5\% & 87.8\% & 17.6\% \\
    2M Page  & 99.8\% & 99.0\% & 99.6\% & 89.6\% \\
    \bottomrule
  \end{tabular*}
  \vspace{-0.5cm}
\end{table}

Secondly, to mitigate the fragmentation, heap managers will kick in and share
the page among threads. Since a page can reside on only one NUMA node, this
creates a phenomenon which we call ``false page-sharing'', where although each
thread instructs the heap manager the requested memory block shall be local, at
least one thread will get remote blocks, which causes unnecessary remote memory
access. In case of automatic page migration, false page-sharing can cause
significant performance degradation like those in cache false-sharing, where the
page is constantly migrated back and forth among accessing threads.

Thirdly, for real-world applications, simple approaches such as first-touch can
be inviable. For example, in multi-block applications\cite{Guo2012A} and
algebraic multigrid methods\cite{Falgout2002H}, the first thread writing to a page are
often not the most frequent writer and first-touch simply would not work.  The
only option is to explicitly mark memory location with the numactl API, which can
be extremely costly as indicated in \cref{sec:exp}. Even in cases where
first-touch do work, due to the complexity of data access patterns, automatic
page migration can often make wrong decisions about what page shall be moved and
causes unnecessary migration overheads as in our experiments in \cref{sec:exp}.

To address the pitfalls above, applications would retreat to carefully arranging
their memory access or taking control of heap management in an
application-specific way, which would soon become an unacceptable burden on
application developers, thus hurt the ease-of-use. These facts motivate us to
seek a reliable yet general way for NUMA-awareness in multi-threaded scientific
applications.

%

\section{Partitioned Shared Memory for NUMA-awareness of Multi-threaded Applications}%
\label{sec:psm}

We propose a ``partitioned shared memory'' approach as a remedy to the pitfalls
of OS-based NUMA-awareness approaches. This approach is a co-design among
applications, runtime system and the underlying OS.\@

\begin{figure}[htpb]
  \vspace{-0.5cm}
  \centering
  \subfloat[Accessing globally shared memory]{%
  \includegraphics[height=3cm]{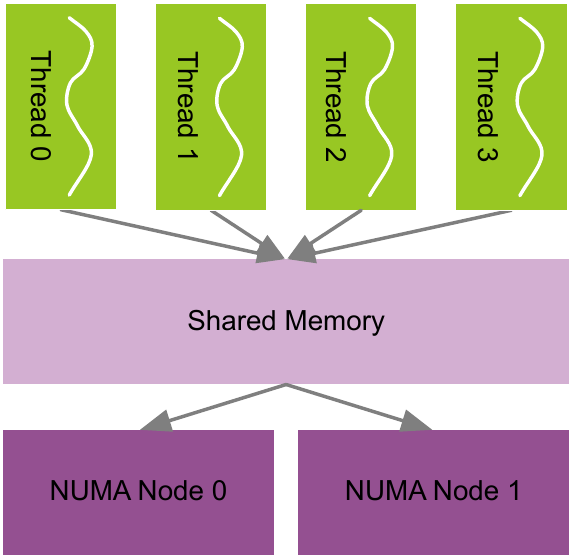}}
  \hspace{0.5cm}
  \subfloat[Accessing partitioned shared memory]{%
  \includegraphics[height=3cm]{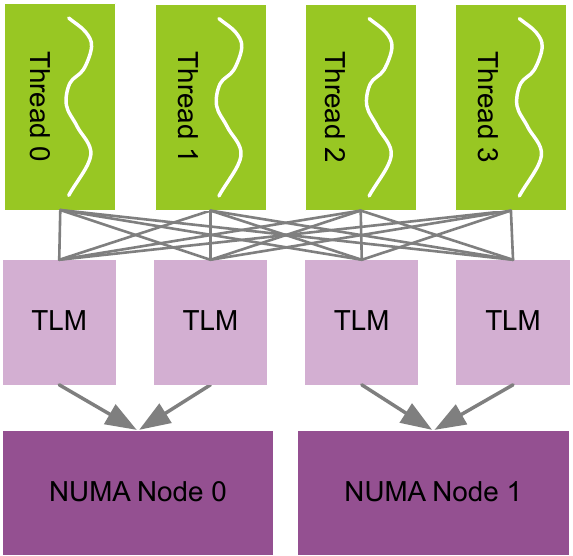}}
  \hspace{0.5cm}
  \subfloat[Software stack overview]{%
  \label{fig:partitioned-shared-memory:c}
  \includegraphics[height=3cm]{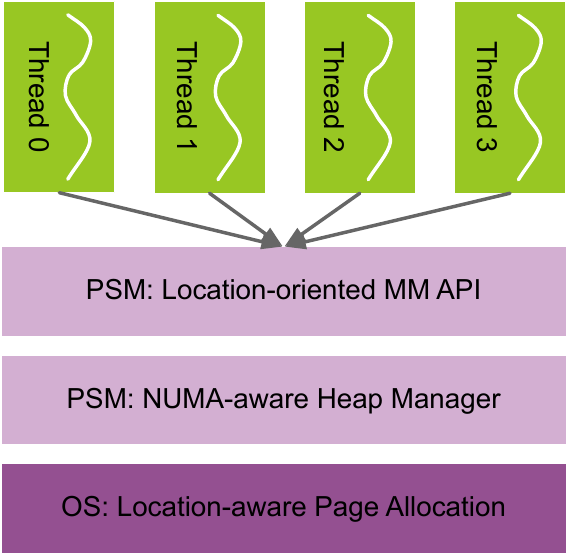}}
  \caption{Illustration of the partitioned shared memory approach: threads
  partition the shared memory and each thread's local part is bind to its
  running NUMA node.}%
  \label{fig:partitioned-shared-memory}
  \vspace{-0.5cm}
\end{figure}

The key idea of partitioned shared memory is illustrated in
\cref{fig:partitioned-shared-memory}. Similar to the idea of ``partitioned
global address space'' idea in parallel languages such as UPC\cite{El2016U},
threads on a computing node partition the OS-provided global view of node's
memory into several thread-local regions (thread local memory, or TLM), and each
thread's local memory is bind to the thread's running NUMA node. This
abstraction provides the application the mechanism to achieve full
NUMA-awareness without the limits of OS-based ones.  For applications using
``owner-compute'' rule, the initializing thread simply allocates data blocks in
the owner thread's local memory. This is similar to first-touch but much more
flexible, since the owner does not need to be the first writer now. This
directly enables NUMA-awareness in applications with complex initialization such
as multi-block applications and algebraic multigrid methods where
``owner-compute'' still holds. Even for those do not use ``owner-compute'', as
long as one can infer which thread would be the most frequent consumer of a data
block as in almost all scientific applications, NUMA-awareness can still be
achieved by allocating the memory in that consumer's local memory. Thus the
partitioned share memory approach enables full NUMA-awareness in multi-threaded
scientific applications.

The partitioned shared memory software stack as illustrated in
\cref{fig:partitioned-shared-memory:c} addresses other pitfalls of OS-based
approaches. Firstly, unlike OS-based approaches, it works at the runtime level
and serves memory as variable-sized blocks instead of fix-sized pages, thus
reduces fragmentation. Secondly, since pages can be shared among threads on the
same NUMA node but never shared among threads on different NUMA nodes, false
page-sharing is eliminated while keeping the fragmentation low. Thirdly, this
approach remains easy-to-use besides the ability to achieve full NUMA-awareness
in complex real-world applications, since the only changes to application code
are memory allocation calls where the allocator shall supply on which thread
shall the memory block resides. Lastly, since this approach only requires the OS
to ensure a page is bind to the specified NUMA node, it is portable.

The key ingredient of partitioned shared memory is that the heap management
shall be NUMA-aware and shall be able to handle allocation requests with
location constraints, which will be addressed by our JArena heap manager in
\cref{sec:jarena-algorithms}.

%

\section{JArena: A NUMA-aware Multi-threaded Heap Manager}%
\label{sec:jarena-algorithms}

To support the partitioned shared memory abstraction, as well as to eliminate false page-sharing,
reduce fragmentation and mitigate overhead caused by page allocation and
binding, we design and develop a NUMA-aware multi-threaded heap manager JArena.
JArena provides two fundamental memory management APIs:
\begin{enumerate}
  \item \emph{location-aware allocation}: \texttt{void* psm\_alloc(size\_t bytes,
    int owner)}
  \item \emph{location-free deallocation}: \texttt{void psm\_free(void* p)}
\end{enumerate}

The challenges for JArena heap manager come from three
aspects: firstly, the manager shall always return blocks from correct locations;
secondly, the manager shall eliminate false page-sharing and reduce
fragmentation; thirdly, the implementation shall be thread-safe and scalable
with tens to hundreds of cores.

\subsection{NUMA-aware and multithread-optimized design of JArena}

JArena stems from the TCMalloc heap manager\cite{Ghemawat2009T}. TCMalloc is a
heap manager optimized for multicore systems, with the ability of low fragmentation
from the carefully-designed segregated storage scheme. To improve scalability,
TCMalloc introduces per-core block caches and a two-level block management
scheme consisting of ``thread cache + central free list''. However, TCMalloc is
not NUMA-aware and suffers from both false page-sharing and returning remote
blocks.

\begin{figure}[htbp]
  \vspace{-0.5cm}
  \centering\includegraphics[width=0.9\textwidth]{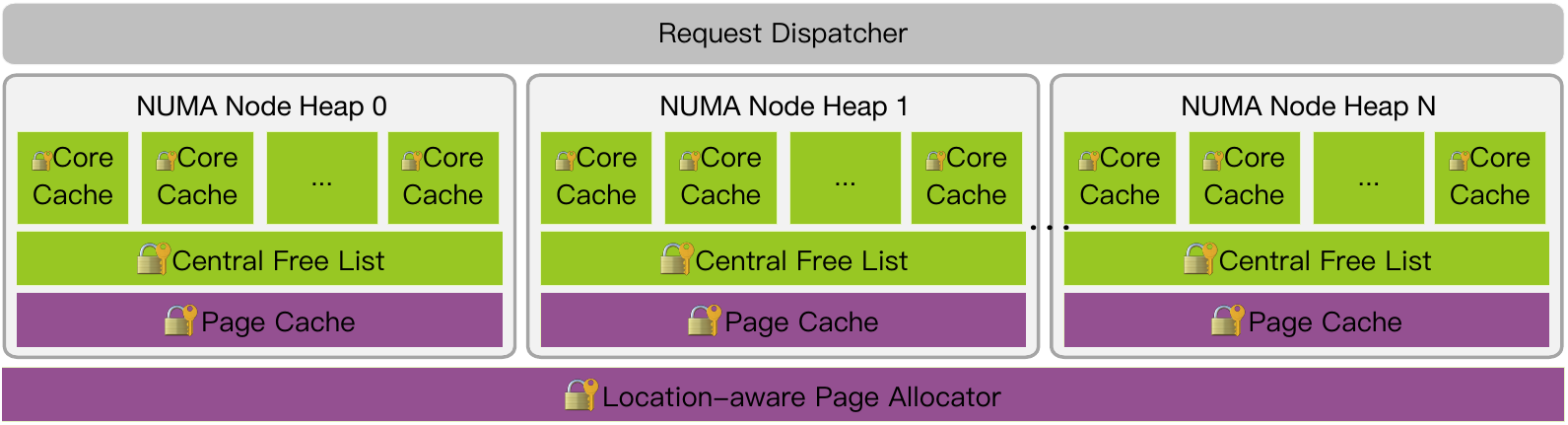}
  \caption{Design the JArena heap manager: the heap is partitioned among
  NUMA nodes and managed seperatedly and the locks are localized.}%
  \label{fig:jarena-arch}
  \vspace{-0.5cm}
\end{figure}

We extend the TCMalloc design to that in \cref{fig:jarena-arch}. Instead of
treating the heap as a whole, we divide it into several independent NUMA node
heaps, each manages memory blocks belonging to one NUMA node in the same way as
TCMalloc. Threads are binding to cores and the \texttt{pcm\_alloc(bytes, tid)}
call will be satisfied by the heap on whose NUMA node \texttt{tid} resides.
This design eliminates false page-sharing, and reduces fragmentation by making
use of the advanced segregated storage scheme in TCMalloc, thus achieves
NUMA-awareness. This design also improves the scalability, since except for the
location-aware page allocator which allocates pages on a certain NUMA node, all
locks are local to a NUMA node heap. Together with the multicore optimization of
TCMalloc, this design can scale to tens to hundreds of cores.

\subsection{Location-aware block serving and recycling}

With the NUMA-aware design, the remaining problem is how to ensure the blocks
are served correctly with the multi-level block caching. This property is
ensured by the location-aware block serving and recycling algorithm, which is an
extension of corresponding algorithms in TCMalloc\cite{Ghemawat2009T}.

To serve an allocation request \texttt{psm\_alloc(bytes, tid)}, JArena first
identifies the core on which thread \texttt{tid} resides, then locks and asks
the corresponding core cache to serve the block. If the core cache can not serve
the block, it will lock the connected central free list and ask for a run of
blocks matching the requested size, who will in turn lock and asks the page
allocator to allocate and bind new pages if not able to fulfill that request.
Since many threads may request blocks from the same thread, the lock on core
cache are mandatory. But in case where these concurrent requests are sparse,
these locks shall not have a significant impact on performance.

To recycle a memory block, JArena first queries all NUMA node heaps for who owns
the block, which is done by checking the address against a two-level page map in
``Page Cache''. Then if the block belongs to the same NUMA node heap the
requesting thread resides, the block is simply returned to the corresponding
core cache for recycling. Otherwise, the block is returned to the central free
list of the owning NUMA node heap for recycling. Further recycling is done by
each NUMA node heap using the same algorithm as TCMalloc.

With the above location-aware block serving and recycling and the above
NUMA-aware multithread-optimized design, JArena addresses the challenges as in
the beginning of this section.

%

\section{Experimental Evaluation}%
\label{sec:exp}

In this section, we evaluate the performance of the proposed partitioned shared
memory NUMA-awareness approach. We first verify that the design and algorithms
in JArena ensures correct location-aware block serving.  Then we compare our
partitioned shared memory against the widely used first-touch approach with
real-world multi-threaded scientific applications.

Details of our experimental platform is shown in \cref{tbl:exp-platform}. This
platform exhibits similar NUMA effect to modern manycore processors with a
cc-NUMA architecture with 32 NUMA nodes. To bind threads and avoid performance
jitters, all threads are bind the CPU cores with \texttt{KMP\_AFFINITY=compact}
in all our experiments. Thus thread $i$ is always bind to core $i$ and NUMA node
$i/8$.
\begin{table}[htbp]
  \vspace{-0.5cm}
  \caption{Details of the experimental platform (a 256-core cc-NUMA node).}%
  \label{tbl:exp-platform}
  \centering
  \begin{tabular*}{\textwidth}{@{\extracolsep{\fill}}rl}
    \toprule
    Processor	      & Intel Xeon 7550 (8 Cores)		     \\
    Number of Processors  & 32 Sockets				      \\
    Memory & DDR3 1600, 32GB on each socket, 1TB in total \\
    Number of NUMA Nodes & 32					       \\
    NUMA Interconnect	& SGI NumaLink 5				 \\
    Min/Max NUMA Distance & 1.0/6.8 (reported by the OS)		     \\
    OS Kernel	      & Linux 2.6.32 \\
    \bottomrule
  \end{tabular*}
  \vspace{-1cm}
\end{table}

\subsection{Verification of NUMA-awareness in JArena}

We design a synthetic benchmark to check if heap managers ensure local
allocation as \cref{lst:jarena-verify-bench}. To emulate the complex but
common case where thread other than the owner frees the memory (for examples
modern C++ applications using smart pointers), each thread will allocate 64 1MB
block, write to them, and free those of the left neighbour. The benchmark
kernel is run once to warm-up the heap manager, then repeated 5 times, where
remote pages are checked using the \texttt{get\_mempolicy} system call or the
page writing time is measured. We want to check if the heap manager is able to
retain memory pages as local. Obviously, a NUMA-aware heap manager shall
contains zero remote pages. We compare JArena with the default heap manager in
GLIBC (namely ptmalloc2) and TCMalloc.
\begin{lstlisting}[
  language=c++,
  label={lst:jarena-verify-bench},
  caption={Benchmark to verify the NUMA-awareness of heap managers}]
char* a[NTHREADS][64];
#pragma omp parallel num_threads(NTHREADS)
{
  int tid = omp_get_thread_num();
  for (int i = 0; i < 64; i++)
    a[tid][i] = alloc_memory_at(1024*1024, tid);
  // check location of pages to verify correctness
  // ...get_mempolicy(b, n)...
  // write to the page to verify writing performance.
  for (int i = 0; i < 64; i++)
    memset(a[tid][i], '1', 1024*1024);
#pragma omp barrier
  for (int i = 0; i < 64; i++)
    free_memory(a[(tid - 1 + NTHREADS) % NTHREADS]);
}
\end{lstlisting}

The remote page counts for for GLIBC, TCMalloc and JArena are shown in
\cref{tbl:remote-page-counts}. It indicates that JArena correctly returns local pages and
is NUMA-aware while TCMalloc returns more remote pages with the increase of
threads due to its NUMA-unawareness and block caching design. It is surprising
that GLIBC returns several remote pages, since for 1MB memory block, it
retreats to system \texttt{mmap} and is actually first-touch. It
reveals that spurious remote page allocation can happen on such cc-NUMA systems
even when first-touch is used.
\begin{table}
  \vspace{-0.5cm}
  \centering
  \caption{Remote page count with different heap managers.}%
  \label{tbl:remote-page-counts}
  \begin{tabular*}{\textwidth}{@{\extracolsep{\fill}}rrrrrrrr}
    \toprule
    Threads  & 8 & 16     & 32     & 64	& 128	  & 192     & 256  \\ \midrule
    GLIBC	 & 0 & 0      & 5      & 389	& 1047	  & 1962    & 2317 \\ TCMalloc
    & 0 & 112079 & 323038 & 779228 & 1684069 & 2598901 & N.A. \\ JArena   & 0 & 0
       & 0	    & 0      & 0       & 0	 & 0	\\ \bottomrule
  \end{tabular*}
  \vspace{-0.5cm}
\end{table}

The page writing time as presented in \cref{tbl:memory-writing-time} shows a
similar trend. Memory blocks from JArena take the least time to write since
pages are local, while those from TCMalloc take 4 times more time since there
are many remote pages. Those from GLIBC take surprisingly long time (100 times
more with 256 threads) because the page allocation under the scene dominates the
overall time. This fact indicates that naive first-touch or memory policy can be
extremely costly on such cc-NUMA systems and block-caching heap managers are
necessary.
\begin{table}
  \vspace{-0.5cm}
  \centering
  \caption{Accumulated memory writing time (in seconds).}%
  \label{tbl:memory-writing-time}
  \begin{tabular*}{\textwidth}{@{\extracolsep{\fill}}rrrrrrrr}
    \toprule
    Threads  & 8	 & 16	 & 32	 & 64	 & 128	 & 192	 & 256	 \\ \midrule
    GLIBC	 & 0.052 & 0.227 & 1.039 & 1.907 & 5.121 & 7.957 & 11.48 \\ TCMalloc &
    0.051 & 0.059 & 0.181 & 0.336 & 0.452 & 0.407 & N.A.  \\ JArena   & 0.039 &
    0.035 & 0.041 & 0.053 & 0.078 & 0.113 & 0.134 \\ \bottomrule
  \end{tabular*}
  \vspace{-0.5cm}
\end{table}

\subsection{Application performance with partitioned shared memory}

We now test the effectiveness of the proposed partitioned shared memory
approach with real-world multi-threaded scientific applications. We choose two
applications: a 2D/3D linear advection application sample\cite{Mo2010}
and an electromagnetic package JEMS-FDTD\cite{Li2014} with FDTD
scheme. Both applications are memory-bounded and adhere to the
``owner-compute'' rule and static load balancing, thus make a good candidate to
compare the performance of first-touch and partitioned shared memory. We modify
the memory allocation in these applications so either the owner thread of a
patch first-touches the memory or make use of \texttt{psm\_alloc} to
allocate memory locally.

We first show result of the linear advection application sample in
\cref{tbl:kernel-time-linadv}, where the wall time of numerical computing for 100 time
steps is aggregated. It concludes that with the increase of NUMA nodes, the
application performs increasingly better than first-touch. JArena starts to
gain advantage at 64 threads and achieves maximum improvements of 3.3 and 2.8
folds for 2D/3D cases respectively. Further investigation shows that in case
where multiple halo-exchange and computation locksteps are involved, under
first-touch, the OS either places pages in remote NUMA nodes or migrates pages
at times, thus it is necessary for approaches such as partitioned shared memory
proposed in this paper.
\begin{table}[htbp]
  \vspace{-0.5cm}
  \centering
  \caption{Accumulated kernel time (in seconds) with different NUMA-awareness approaches
    for the linear advection application sample.}%
  \label{tbl:kernel-time-linadv}
  \begin{tabular*}{\textwidth}{@{\extracolsep{\fill}}rrrrrrrrrrrrr}
    \toprule
    & \multicolumn{6}{c}{2D Linear Advection} & \multicolumn{6}{c}{3D Linear Advection} \\
    \cmidrule{2-7} \cmidrule{8-13}
    Threads     & 8    & 16   & 32	 & 64	& 128  & 256
                & 8    & 16   & 32   & 64 & 128 & 256 \\
    \midrule
    First Touch & 89.6 & 44.8 & 23.7 & 16.0 & 11.9 & 17.7
                & 59.6 & 29.8 & 15.6 & 10.6 & 6.9 & 9.1 \\
    JArena      & 90.4 & 45.2 & 22.7 & 11.2 & 5.6  & 4.1
                & 60.1 & 30.1 & 15.1 & 7.5  & 3.8 & 2.4 \\
    Improvements & -0.01 & -0.01 & 0.04 & 0.43 & 1.1 & 3.3
                 & -0.01 & -0.01 & 0.03 & 0.41 & 0.82 & 2.8 \\
    \bottomrule
  \end{tabular*}
  \vspace{-0.5cm}
\end{table}

We then show results of JEMS-FDTD solving a plane-wave scattering problem with
100 time steps in \cref{tbl:jemstd-kernel-time}. The data show similar trends
as of the linear advection application sample, where JArena performs
increasingly better than first-touch with the increase of NUMA nodes. JArena
starts to gain advantage at 64 threads and achieves maximum improvements of 4.3
folds at 256 threads. The spurious increase of computation with 256 threads
needs further investigation, but may be caused by overhead in the cc-NUMA
protocols.
\begin{table}[htbp]
  \vspace{-0.5cm}
  \centering
  \caption{Accumulated kernel time (in seconds) with different NUMA-awareness approaches
    for JEMS-FDTD application.}%
  \label{tbl:jemstd-kernel-time}
  \begin{tabular*}{\textwidth}{@{\extracolsep{\fill}}rrrrrrr}
    \toprule
    Threads     & 8    & 16   & 32	 & 64  & 128 & 256  \\
    \midrule
    First Touch & 47.5 & 23.7 & 12.4 & 7.3 & 8.4 & 28.1 \\
    JArena      & 46.8 & 23.3 & 12.0 & 6.4 & 4.2 & 5.3  \\
    Improvements & 0.01 & 0.02 & 0.03 & 0.14 & 1.0 & 4.3 \\
    \bottomrule
  \end{tabular*}
  \vspace{-0.2cm}
\end{table}

%

\section{Related Works}%
\label{sec:related-works}

First-touch has long been recommended as the major approach for multi-threaded
scientific applications as suggested in~\cite{Rabenseifner2009, Hager2010} and
practiced by applications such as stencils\cite{Shaheen2012}. Beyond
first-touch, advanced page placements such as the numactl API\cite{Kleen2005}
and automatic page migration\cite{Goglin2009E} and balancing\cite{Gorman2012A}
are introduced in the Linux kernel. However, these OS-based approaches can
suffer from false page-sharing and fragmentation, as well as imperfect OS
optimization as suggested by this paper and in algorithmic skeletons
by~\cite{Metzger2018N}.

To hide the complexity of NUMA while optimizing the performance, runtime-based
approaches are widely exploited, such as NUMA-aware OpenMP\cite{Broquedis2010F}
and TBB\cite{Majo2015}. Task-based runtimes and frameworks also make extensive
NUMA-aware optimizations as those in PaRSEC\cite{Bosilca2013},
Charm++\cite{Pilla2010} and Boxlib\cite{Zhang2016B}.  Except for domain-specific
frameworks such as Boxlib and Uintah, these optimizations can not achieve full
NUMA-awareness in complex applications since they rely on assumptions of
applications' data access patterns. And domain-specific frameworks can be
restricted in their application domains.

Similar to the co-design idea in this paper, Pousa Ribeiro et.al.\ proposed
Minas to allow application to explicitly manage or automatically tune their
memory affinity in~\cite{PousaRibeiro2011I}. Their approach does not introduce
the partitioned shared memory abstraction thus leaves all complexity to the
application. Other approaches based on NUMA-aware heap managers include a
NUMA-aware TCMalloc~\cite{Kaminski2009} and memkind~\cite{Cantalupo2015M}, which
also fall short of application-level conceptual abstractions such as partitioned
shared memory. The partitioned shared memory approach borrows from the PGAS idea
in parallel languages such as UPC\cite{El2016U}.

%

\section{Conclusion and Outlook}%
\label{sec:conclusion}

To address the drawbacks of traditional OS-based NUMA-awareness approaches and
achieve full NUMA-awareness in multi-threaded parallel scientific applications,
we propose a partitioned shared memory approach in this paper, which consists of
an application-level conceptual abstraction of partitioned shared memory and a
companying NUMA-aware multi-threaded heap manager named JArena. The partitioned
shared memory abstraction enables parallel applications to explicitly reason
which thread shall own the allocated block, thus maximizes their ability to
achieve NUMA-awareness and preserve locality. JArena then ensures allocated
blocks are pinned to the owner's local memory. JArena manages memory at the
block granularity which eliminates false page-sharing and minimizes
fragmentation. Owing to the full NUMA-awareness possibility of this approach,
real-world multi-threaded scientific applications such as JEMS-FDTD can improve
up to 4.3 folds the performance on a 256-core cc-NUMA system.

One shortcoming of the proposed approach is that the application shall label
explicitly which thread shall own the block to allocate. We would like to
investigate the possibility to infer this information with the help of a
task-based runtime such as Legion, StarPU, and PaRSEC\@. We also plan to
optimize the heap manager algorithms for better performance and scalability.

\section*{Acknowledgments}

The authors would like to thank Dr. Linping WU from High Performance Computing
Center of Institute of Applied Physics and Computational Mathematics for his help
on understanding the OS interferences on cc-NUMA systems.  Dr. Xu LIU and Dr.
Xiaowen XU contribute several key ideas to the refinement of this paper.

%
\renewcommand{\bibsection}{\section*{References}}
\bibliographystyle{splncs04}
\begingroup
\small 
\bibliography{references}
\endgroup

\end{document}